\documentclass[prd,preprintnumbers,amsmath,amssymb,superscriptaddress,nofootinbib]{revtex4}

\usepackage{graphicx}
\usepackage{dcolumn}
\usepackage{bm}
\usepackage{color}
\usepackage{caption}
\usepackage{slashed}

\newcommand{\nn}{\nonumber}

\def\beq{\begin{equation}}
\def\eeq{\end{equation}}
\def\bqa{\begin{eqnarray}}
\def\eqa{\end{eqnarray}}

\def\dfrac{\displaystyle\frac}

\allowdisplaybreaks[2]

\begin{document}

\title{Soft pattern of Rutherford scattering from heavy target mass expansion}

\author{Yu Jia\footnote{jiay@ihep.ac.cn}}
\affiliation{Institute of High Energy Physics and Theoretical Physics Center for Science Facilities, Chinese Academy of Sciences, Beijing 100049, China\vspace{0.2cm}}
\affiliation{School of Physics, University of Chinese Academy of Sciences, Beijing 100049, China\vspace{0.2cm}}

\author{Jia-Yue Zhang\footnote{zhangjiayue@ihep.ac.cn}}
\affiliation{Institute of High Energy Physics and Theoretical Physics Center for Science Facilities, Chinese Academy of Sciences, Beijing 100049, China\vspace{0.2cm}}
\affiliation{School of Physics, University of Chinese Academy of Sciences, Beijing 100049, China\vspace{0.2cm}}

\date{\today}

\begin{abstract}
We investigate the soft behavior of the tree-level Rutherford scattering process.
We consider two types of Rutherford scattering, a low-energy massless point-like projectile (say, a spin-${1\over 2}$ or spin-$0$ electron) to hit
a static massive composite target particle carrying various spins (up to spin-$2$), and a slowly-moving light projectile hits a heavy static composite target.
For the first type, the unpolarized cross sections in the laboratory frame are found to exhibit universal forms
in the first two orders of $1/M$ expansion, yet differ at the next-to-next-to-leading order (though some terms at this order
still remain to be universal or depend on the target spin in a definite manner). For the second type, at the lowest order in electron velocity expansion, 
through all orders in $1/M$, the unpolarized cross section is universal (also not sensitive to the projectile spin). The universality partially breaks down
at relative order-$v^2/M^2$, though some terms at this order are still universal or depend on the target spin in a specific manner. 
We also employ the effective field theory approach to reproduce the soft behavior of the differential cross sections
for the target particle being a composite Dirac fermion.
\end{abstract}

\maketitle

\section{Introduction}

Rutherford scattering is one of the most classic experiments in the history of physics. Originally Gegier and Marsden bombed the nonrelativistic
the $\alpha$ particle beam on the gold foil in $1909$~\cite{gegier1909diffuse}.  Shortly after, in 1911 Rutherford introduced the revolutionary
concept of atomic nucleus, and successfully explained the experimental results by simply exploiting classical mechanics~\cite{Rutherford:1911zz}.
Without exaggeration, Rutherford scattering experiment heralded the advent of nuclear physics and quantum mechanics.

Half a century later, a new form of Rutherford scattering experiments conducted at SLAC,
{\it i.e.}, bombing an energetic electron beam onto the fixed nucleus target,
played a pivotal role in unravelling the internal structure of a nucleon.
Through $ep$ elastic scattering experiments, the electromagnetic form factors of a proton have been measured over a large range of $Q^2$.
From their profiles at lower $Q^2$ end, one can infer the proton's gross features such as the charge radius and magnetic dipole.
It is interesting to note that, there exists a decade-long puzzle about the proton's charge radius, {\it e.g.},
the five standard discrepancy between the value extracted from the $ep$ elastic scattering {and ordinary hydrogen spectra} and from the
muonic hydrogen Lamb shift measurement~\cite{Mohr:2012tt,Pohl:2010zza}.

To infer the gross feature of composite nuclei from Rutherford scattering, the exchanged virtual photon necessarily carries the long wavelength
(hence low resolution). To this purpose, it is appropriate to concentrate on the low-energy behavior of the Rutherford scattering process.
It is worth mentioning that, another basic QED process, Compton scattering, in which a photon beam shining on a composite spinning target particle,
can also be used to probe the internal structure of the atomic nuclei. The soft behavior of the angular distribution of the Compton scattering
in the laboratory frame has been thoroughly studied by Gell-Mann and Low in the 1960s~\cite{Low:1954kd,Gell-Mann:1954wra},
which turns out to possess some simple and universal structure.
Based on the intuitive multipole expansion picture, one naturally anticipates that the soft limit of Rutherford scattering may also
exhibit some universal and simple patterns.

It is the goal of this work to comprehensively investigate the soft behavior of the two typical types of Rutherford scattering, {\it i.e.},
a low-energy massless/a slowly-moving light projectile hits a static, heavy, composite spinning target particle.
For simplicity, we assume the projectile to be a structureless point particle, say, the spin-${1\over 2}$ or spin-$0$ electron.
For concreteness, we choose the spin of the composite target particle to range from 0 to 2. We find in both cases, the differential cross section  of
Rutherford scattering exhibit the universal behavior in the first two terms upon heavy target mass expansion, yet differ at the next-to-next-to-leading
order (depending on target spin). We conjecture this pattern may persist for the heavy target particle with arbitrary spin.

The rest of the paper is structured as follows.
In section~\ref{sec:general:amplitude}, we present the expression of the tree-level
Rutherford scattering amplitude involving a heavy composite spinning target particle, specifying the
parametrization of the electromagnetic form factors of the target particle.
Section~\ref{sec:soft:behavior} is the main body of the paper, where we present the soft behavior of two types of Rutherford scattering
cross section up to next-to-next-to-leading order in heavy target mass expansion, assuming the projectile is the point spin-${1\over 2}$
electron. We explicitly demonstrate the universal behavior of the first two terms upon heavy target mass expansion, and the difference at
NNLO.
In section~\ref{EFT:interpretation}, we attempt to apply the heavy particle effective theory (HPET) and nonrelativistic QED (NRQED)
to reproduce the soft behavior for the case of a spin-${1\over 2}$ target particle.
We summarize in section~\ref{sec:summary}.
In Appendix, we also demonstrate the main conclusion still holds once the projectile is replaced by a point-like spinless electron.

\section{Amplitude of Rutherford scattering involving a heavy composite target particle}
\label{sec:general:amplitude}

\begin{figure}[t]
\center{
\includegraphics[scale=0.7]{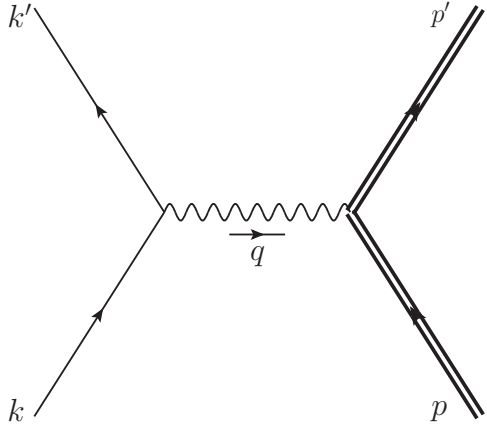}
\caption {\label{fig:feynman diag} Tree-level Feynman diagram for Rutherford scattering process $e N\to e N$.
}}
\end{figure}

To be specific, let us consider the Rutherford scattering process $e(k) N(p)\to e(k^\prime) N(p^\prime)$, where $N$ represents a heavy target particle.
At tree-level, Rutherford scattering is induced by a single photon $t$-channel exchange, as depicted in Fig.~\ref{fig:feynman diag}.
The scattering amplitude can be written as
\begin{align}
  \mathcal{M}=\frac{e^2 g_{\mu\nu}}{q^2}\langle e^-\left(k'\right)
  \vert J^\mu\vert e^-\left(k\right)\rangle\langle N\left(p',\lambda'\right)|J^\nu|N\left(p,\lambda\right)\rangle,
\label{eq: amplitude}
\end{align}
where $J^\mu$ denotes the electromagnetic current,  $q=k-k'$ represents the momentum exchange due to the virtual photon.
$\lambda,\lambda'$ denote the polarization indices for the massive spinning target particle.
For simplicity, we have suppressed the spin index of the electron, and also neglected the electron mass.

The electromagnetic transition matrix element involving nucleus in \eqref{eq: amplitude} is generally a nonperturbative object, since the heavy target $N$ is assumed to be
any massive composite particle. However, this matrix element can be generally decomposed into the linear combination of
independent electromagnetic form factors (FFs) according to Lorentz group representation~\cite{Cotogno:2019vjb}:
\begin{subequations}
  \begin{align}
\langle N\left(p',\lambda'\right)|J^\nu|N\left(p,\lambda\right)\rangle_{s=0}=&2P^\mu F_{1,0}\left(\dfrac{q^2}{M^2}\right),\\
\langle N\left(p',\lambda'\right)|J^\nu|N\left(p,\lambda\right)\rangle_{s=\frac{1}{2}}=&\bar{u}(p',\lambda')\left[2P^\mu F_{1,0}\left(\dfrac{q^2}{M^2}\right)+\mathrm{i}\sigma^{\mu\nu}q_\nu F_{2,0}\left(\dfrac{q^2}{M^2}\right)\right]u(p,\lambda),\\
\langle N\left(p',\lambda'\right)|J^\nu|N\left(p,\lambda\right)\rangle_{s=1}=&-\varepsilon^*_{\alpha'}(p',\lambda')
  \bigg\{2P^ \mu\left[g^{\alpha'\alpha}F_{1,0}\left(\dfrac{q^2}{M^2}\right)-\dfrac{q^{\alpha'}q^\alpha}{{2M^2}}F_{1,1}\left(\dfrac{q^2}{M^2}\right)\right]\nn\\
  &-\left(g^{\mu\alpha'}q^{\alpha}-g^{\mu\alpha}q^{\alpha'}\right)F_{2,0}\left(\dfrac{q^2}{M^2}\right)\bigg\}\varepsilon_{\alpha}(p,\lambda),\\
\langle N\left(p',\lambda'\right)|J^\nu|N\left(p,\lambda\right)\rangle_{s=\frac{3}{2}}=&-\bar{u}_{\alpha'}(p',\lambda')
\bigg\{2P^ \mu\left[g^{\alpha'\alpha}F_{1,0}\left(\dfrac{q^2}{M^2}\right)-\dfrac{q^{\alpha'}q^\alpha}{{2M^2}}F_{1,1}\left(\dfrac{q^2}{M^2}\right)\right]\nn\\
&+\mathrm{i}\sigma^{\mu\nu}q_\nu\left[g^{\alpha'\alpha}F_{2,0}\left(\dfrac{q^2}{M^2}\right)-\dfrac{q^{\alpha'}q^\alpha}{{2M^2}}F_{2,1}\left(\dfrac{q^2}{M^2}\right)\right]\bigg\}u_{\alpha}(p,\lambda),\\
\langle N\left(p',\lambda'\right)|J^\nu|N\left(p,\lambda\right)\rangle_{s=2}=&\varepsilon^*_{\alpha'_1\alpha'_2}(p',\lambda')\bigg\{2P^\mu\bigg[g^{\alpha'_1\alpha_1}g^{\alpha'_2\alpha_2}F_{1,0}\left(\dfrac{q^2}{M^2}\right)-\dfrac{q^{\alpha'_1}q^{\alpha_1}}{2M^2}g^{\alpha'_2\alpha_2}F_{1,1}\left(\dfrac{q^2}{M^2}\right)\nn\\
&+\dfrac{q^{\alpha'_1}q^{\alpha_1}}{2M^2}\dfrac{q^{\alpha'_2}q^{\alpha_2}}{2M^2}F_{1,2}\left(\dfrac{q^2}{M^2}\right)\bigg]-\left(g^{\mu\alpha'_2}q^{\alpha_2}-g^{\mu\alpha_2}q^{\alpha'_2}\right)\nn\\
&\times\bigg[g^{\alpha'_1\alpha_1}F_{2,0}\left(\dfrac{q^2}{M^2}\right)-\dfrac{q^{\alpha'_1}q^{\alpha_1}}{2M^2}F_{2,1}
\left(\dfrac{q^2}{M^2}\right)\bigg]\bigg\}\varepsilon_{\alpha_1\alpha_2}(p,\lambda).
\end{align}
\label{eq:FF}
\end{subequations}
The various electromagnetic FFs are normalized to be dimensionless.
$P=(p+p')/2$ is the average momentum of the target particle, and $M$ is the mass of target particle.
$u$, $\varepsilon^\mu$, $u^\mu$, $\varepsilon^{\alpha\beta}$
denote the wave function for the spin-${1\over 2}$, 1, ${3\over 2}$, and 2 particles, respectively.
Only keeping those Lorentz structures that obey the current conservation,
one observes that the number of independent electromagnetic FFs is $2s+1$ for
target particle with spin $s$.
Note that the decomposition of the electromagnetic transition matrix element involving charged particle carrying various spin
has been widely studied~\cite{Scadron:1968zz,Williams:1970ms}.

The electromagnetic FFs in \eqref{eq:FF} encode the internal structure of the composite target particle.
In principle, they can be extracted from experiments or computed by nonperturbative theoretical tools. Although the concrete profiles of
various FFs depend on the specific target particle, their values near the zero momentum transfer do characterize the electromagnetic multipole
moments of the composite target particle.
For example, $F_{1,0}(0)=Z$ denotes the total electric charge of the target particle in units of $e$.
 $F_{1,0}(0)+F_{1,1}(0)$, $F_{2,0}(0)$ and $F_{2,0}(0)+F_{2,1}(0)$ are the electric quadrupole moment, magnetic dipole moment and magnetic octupole moment of the composite target particle, in units of ${e\over 2M}$, ${e\over M^2}$, and ${e\over 2M^3}$, respectively~\cite{Nozawa:1990gt}.
It is interesting to note that, the charge radius of a proton,
may also be expressed as $r_p= {3\over 2 M^2} [-{F_{1,0}}(0)+{4F'_{1,0}}(0)+F_{2,0}(0)]$~\footnote{The Taylor expansion of the form factors
around the origin is understood as $F_n (q^2/M^2)= F_n(0)+ F_n'(0){q^2\over M^2}+ {\mathcal O}(1/M^4)$.}.

\section{Low-energy Rutherford scattering in heavy target mass expansion }
\label{sec:soft:behavior}

Squaring the amplitude \eqref{eq: amplitude}, averaging over spins in the initial state and summing over the polarizations in the final states,
One can straightforwardly obtain the unpolarized differential cross sections of Rutheford scattering in the laboratory frame for various target particle species.
In deriving the unpolarized cross sections, the following spin sum relations are useful:
\begin{subequations}
  \begin{align}
    &\sum_{\lambda}u(p,\lambda)\bar{u}(p,\lambda)=\dfrac{\slashed p+M}{2M},\\
    &\sum_{\lambda}\varepsilon_\alpha(p,\lambda)\varepsilon^*_{\alpha'}(p,\lambda)=\eta_{\alpha\alpha'},\\
    &\sum_{\lambda}u_\alpha(p,\lambda)\bar{u}_{\alpha'}(p,\lambda)=-\dfrac{\slashed p+M}{2M}\left({g_{\alpha\alpha'}-\frac{1}{3}\gamma_\alpha\gamma_{\alpha'}-\frac{2p_\alpha p_{\alpha'}}{3M^2}+\frac{\gamma_{\alpha'}p_\alpha-\gamma_{\alpha'}p_{\alpha}}{3M}}\right),\\
    &\sum_{\lambda}\varepsilon_{\alpha_1\alpha_2}(p,\lambda)\varepsilon^*_{\alpha'_1\alpha'_2}(p,\lambda)=\eta_{\alpha_1\alpha'_1}\eta_{\alpha_2\alpha'_2}+\eta_{\alpha_1\alpha'_2}\eta_{\alpha_2\alpha'_1}-\dfrac{2}{3}\eta_{\alpha_1\alpha_2}\eta_{\alpha'_1\alpha'_2},
\end{align}
\label{eq:spin_sum}
\end{subequations}
with $\eta_{\alpha\beta} \equiv -g_{\alpha\beta}+\dfrac{p_\alpha p_\beta}{M^2}$. Note the Dirac
spinor wave function is normalized as $\bar{u}(p,r) u(p,s)=\delta^{rs}$,

\subsection{massless spin-1/2 projectile  }
\label{mass:0:fermion:projectile}

We first consider the modern $ep$ elastic scattering experiment. In such case, the incident electron is treated as massless, and we are concerned with
the low-energy limit $|{\bf k}| \ll M$.

We focus on the Rutherford scattering in the laboratory frame, with the four-momentum of the target particle in the initial state signified by
$p^\mu=(M,{\bf 0})$.
The corresponding differential unpolarized cross section is defined by
\begin{align}
   &\dfrac{\mathrm{d}\sigma}{\mathrm{d}\cos\theta}=
    \dfrac{1}{2 |{\bf k}|}\cdot\dfrac{1}{2M}\cdot\dfrac{{\bf k'}^2}{8\pi|{\bf k}| M}\left(\dfrac{1}{2}\dfrac{1}{2s+1}\sum_\text{spins}\left|\mathcal{M}\right|^2\right),
    \label{def: cross:section:first:Rutherford}
\end{align}
where $\theta$ denotes the polar angle between the incident and the reflected electron.
$|{{\bf k}^\prime}|$ is a function of $|{\bf k}|$, $\cos\theta$ and $M$:
\beq
|{\bf k}'|= {|{\bf k}| \over 1+{|{\bf k}|\over M}\left ( 1-\cos\theta\right)}.
\label{kprime:k:relation}
\eeq

The full expressions of the unpolarized cross sections are generally lengthy and cumbersome-looking,
from which it is difficult to recognize any clear pattern about the dependence on the heavy target particle spin.
Hopefully, once the heavy target mass expansion is conducted, the soft behavior of the Rutherford scattering will become transparent and one
may readily identify some simple pattern.

After expanding both the squared amplitude and the phase space measure (the factor ${{\bf k}'}^2/{\bf k}^2$)
in \eqref{def: cross:section:first:Rutherford} powers of $1/M$,
the differential Rutherford scattering cross sections become much simpler.
We find the first two orders in heavy target expansion
are universal, {\it e.g.}, independent of the heavy target spin:
\begin{align}
   &\dfrac{\mathrm{d}\sigma}{\mathrm{d}\cos\theta}
    =\dfrac{\pi\alpha^2Z^2\cos^2\frac{\theta}{2}}{2 {\bf k}^2\sin^4\left(\frac{\theta}{2}\right)}-\frac{\pi\alpha^2Z^2\cos^2\frac{\theta}{2}}{M |{\bf k}|\sin ^2\left(\frac{\theta}{2}\right)}+\mathcal{O}\left({1\over M^2}\right).
\label{eq: cross section-NLO}
\end{align}
For clarity we have substituted $F^\prime_{1,0}=Z$. This result is intuitively clear,  in the soft limit,
the long wavelength photon can only feel the total charge of the composite target particle,
insensitive to any further details about its internal structure.

In contrast, the next-to-next-to-leading-order (NNLO) terms in heavy target mass expansion
do vary with different heavy target particles:
\begin{subequations}
  \begin{align}
    \left(\dfrac{\mathrm{d}\sigma}{\mathrm{d}\cos\theta}\right)_\text{NNLO}^{s=0}=&
    -\dfrac{4\pi\alpha^2}{M^2\sin^2\frac{\theta}{2}}{\left(F'_{1,0}Z\cos^2\dfrac{\theta}{2}+\dfrac{1}{8}Z^2\cos^2{\theta}-\dfrac{1}{8}Z^2\right)}\\
    \left(\dfrac{\mathrm{d}\sigma}{\mathrm{d}\cos\theta}\right)_\text{NNLO}^{s=\frac{1}{2}}=&
    -\dfrac{4\pi\alpha^2}{M^2\sin^2\frac{\theta}{2}}\bigg[\frac{1}{16}F^2_{2,0}\left(\cos\theta-3\right)\label{eq: cross section-NNLO-1/2}+\dfrac{1}{4}\cos^2\dfrac{\theta}{2}\left(4F'_{1,0}Z+F_{2,0}Z+Z^2\cos\theta-\dfrac{3}{2}Z^2\right)\bigg]\\
    \left(\dfrac{\mathrm{d}\sigma}{\mathrm{d}\cos\theta}\right)_\text{NNLO}^{s=1}=&
    -\dfrac{4\pi\alpha^2}{M^2\sin^2\frac{\theta}{2}}\bigg[\frac{1}{24}F^2_{2,0}\left(\cos\theta-3\right)+\dfrac{1}{4}\cos^2\dfrac{\theta}{2}\left(4F'_{1,0}Z-\dfrac{2}{3}F_{1,1}Z+\dfrac{2}{3}F_{2,0}Z+Z^2\cos\theta-\dfrac{5}{3}Z^2\right)\bigg]\\
    \left(\dfrac{\mathrm{d}\sigma}{\mathrm{d}\cos\theta}\right)_\text{NNLO}^{s=\frac{3}{2}}=&
    -\dfrac{4\pi\alpha^2}{M^2\sin^2\frac{\theta}{2}}\bigg[\frac{5}{144}F^2_{2,0}\left(\cos\theta-3\right)
    +\dfrac{1}{4}\cos^2\dfrac{\theta}{2}\left(4F'_{1,0}Z-\dfrac{2}{3}F_{1,1}Z+F_{2,0}Z+Z^2\cos\theta-\dfrac{13}{6}Z^2\right)\bigg]\\
    \left(\dfrac{\mathrm{d}\sigma}{\mathrm{d}\cos\theta}\right)_\text{NNLO}^{s=2}=&
    -\dfrac{4\pi\alpha^2}{M^2\sin^2\frac{\theta}{2}}\bigg[\frac{1}{32}F^2_{2,0}\left(\cos\theta-3\right)
    +\dfrac{1}{4}\cos^2\dfrac{\theta}{2}\left(4F'_{1,0}Z-\dfrac{2}{3}F_{1,1}Z+\dfrac{2}{3}F_{2,0}Z+Z^2\cos\theta-\dfrac{7}{3}Z^2\right)\bigg]
  \end{align}
  \label{eq: cross section-NNLO}
  \end{subequations}
For notational brevity, we have neglected the argument 0 in various form factors.
We observe that $F'_{1,0}Z$, $Z^2\cos\theta$ and $F_{1,1}Z$ terms with a prefactor $\cos^2(\theta/2)$ are still universal, {\it i.e.},
independent of the target spin. In fact, the $F'_{1,0}Z$ and $Z^2\cos\theta$ terms
actually have the same origin of the LO and NLO cross sections,
which correspond to different terms in Taylor expansion of $F^2_{1,0}(q^2/M^2)$ in the squared LO amplitude and phase space measure.
The coefficient of the $F_{2,0}Z$ term seems to reflect the spin-statistic characteristic of the target particle.
For fermions, the coefficient is $1$, while for bosons $2/3$.

Although the coefficients of $F^2_{2,0}(\cos\theta -3)$ inside the square bracket
depend on the target particle spin $s$, they seem to fit into the expression
${1+s\over 48s}$ (for $s=1/2,1,3/2,2$). It is curious whether this pattern still persists for higher target spin or not.

\subsection{Light non-relativistic spin-1/2 projectile}
\label{light:NR:spin:half:projectile}

Next we turn to the soft limit of the original prototype of Rutherford scattering process, that is, a slowly moving light particle hits a heavy static target.
We again assume the projectile is a Dirac fermion, whose mass and momentum are denoted by $m$ and $\bf k$.

The differential cross section for this type of Rutherford scattering in the laboratory frame is defined by
\begin{align}
\dfrac{\mathrm{d}\sigma}{\mathrm{d}\cos\theta}=\dfrac{1}{32\pi M}\left[{p'^0+k'^0\left(1-\dfrac{|\mathbf{k}|}{|\mathbf{k}'|}\cos\theta\right)}\right]^{-1}\dfrac{|\mathbf{k}'|}
{|\mathbf{k}|}\left|\mathcal{M}\right|^2.
\label{def:diff:cross:section:massless:lab:frame}
\end{align}

The resulting expressions are rather lengthy. Fortunately,we are only interested in its soft behavior. Since
there are three widely separated scales in this process, which obey ${\bf k}\ll m \ll M$,
the appropriate way of extracting the soft behavior is to expand the differential cross sections in powers of
$v=|{\bf k}|/m$ (velocity of the projectile) and $1/M$ simultaneously.
The necessity of performing double expansion renders this case
somewhat more complicated than the preceding case as discussed in section~\ref{mass:0:fermion:projectile}.

Interestingly, in the lowest order in velocity yet to all orders in $1/M$,
the differential cross sections scales as $1/|{\bf k}|^4$,
which takes a uniform form:
\bqa
\left(\dfrac{\mathrm{d}\sigma}{\mathrm{d}\cos\theta}\right)^{s}_{(v^0)}
& = & {2 \pi Z^2 \alpha^2 \over {\bf k}^4} { m^2 (M+m)^2 \left(\sqrt{M^2-m^2\sin^2\theta}+m \cos\theta\right)^2 \over M \sqrt{M^2-m^2\sin^2\theta}\left(M-\cos\theta \sqrt{M^2-m^2\sin^2\theta}+m \sin^2\theta\right)^2}
\nn\\
&= &\frac{8 \pi  Z^2 \alpha ^2 m^2}{\mathbf{k}^4 \sin^4\frac{\theta}{2}}-\frac{\pi Z^2  \alpha ^2  m^4}{M^2\mathbf{k}^4}+
\mathcal{O}\left({m^6\over M^4 {\bf k}^4}\right).
\label{eq: NR-LO}
\eqa

At the next-to-leading order in velocity expansion, the differential cross sections scale
as $1/|{\bf k}|^2$, whose explicit expressions are still rather complicated yet
vary with different target species. Nevertheless, once the heavy target mass expansion is
conducted, some clear pattern emerges:
\begin{align}
\left(\dfrac{\mathrm{d}\sigma}{\mathrm{d}\cos\theta}\right)^{s}_{(v^2)} = &\dfrac{\pi\alpha^2}{\mathbf{k}^2\sin^2\frac{\theta}{2}}
\left[\frac{Z^2 \cos^2\frac{\theta}{2}}{2\sin^2\frac{\theta}{2}}
-\frac{Z^2 m\cos^2\frac{\theta}{2}}{M}
-\frac{Z m^2}{4M^2 } f^s_{\rm NNLO} +
\mathcal{O}\left(\dfrac{1}{M^3}\right)\right],
\end{align}
where
\begin{subequations}
\begin{align}
f_{\rm NNLO}^{s=0}=&16F'_{1,0}+Z\cos\theta-Z,
\\
f_{\rm NNLO}^{s={1\over 2}}=&16F'_{1,0}+Z\cos\theta+4F_{2,0}-3Z,
\label{eq:NR-NNLO:spin-half}
\\
f_{\rm NNLO}^{s=1}=&16F'_{1,0}+Z\cos\theta-\dfrac{8}{3}F_{1,1}+ \dfrac{8}{3}F_{2,0}-\dfrac{11}{3}Z,
\\
f_{\rm NNLO}^{s={3\over 2}}=&16F'_{1,0}+Z\cos\theta-\dfrac{8}{3}F_{1,1}+4F_{2,0}-\dfrac{17}{3}Z,
\\
f_{\rm NNLO}^{s=2}=&
16F'_{1,0}+Z\cos\theta-\dfrac{8}{3}F_{1,1}+\dfrac{8}{3}F_{2,0}-\dfrac{19}{3}Z.
\end{align}
\label{eq:NR-NNLO}
\end{subequations}
The LO and NLO terms in $1/M$ expansion are universal. The NNLO terms begin to exhibit target spin dependence. However,
even at ${\cal O}(v^2/M^2)$,
the $F^\prime_{1,0}$, $F_{1,1}$ and $Z\cos\theta$ terms still seem to be universal, {\it i.e.},  independent of the
target particle spin.  The coefficient of $F_{2,0}$ seems to reflect the spin-statistic characteristic of the target particle.
For fermions, the coefficient is $4$, while for bosons ${8\over 3}$.

\section{Reproducing the soft behavior from effective field theory}
\label{EFT:interpretation}

The low-energy limit of Rutherford scattering is largely dictated by a heavy target particle interacting with a soft photon.
Therefore, it is natural to expect the soft behavior can be reproduced by an effective field theory analogous to heavy quark effective
theory (HQET), which automatically incorporates the heavy target mass expansion.
In this section, we will specialize to the case of a spin-$1/2$ composite target particle.

Originally, HQET is designed to describe a structureless heavy quark interacting with soft gluons~\cite{Eichten:1989zv,Georgi:1990um}.
Due to the asymptotic freedom property of QCD, the Wilson coefficients can be computed in perturbation theory through perturbative matching procedure.

The key idea of HQET can be readily transplanted to the case of a heavy composite particle interacting with a soft photon,
as long as the photon wavelength is too long to deeply probe the internal structure of the composite target.
As a price, one is generally unable to calculate various Wilson coefficients from the top-down perspective.
The internal structure of the composite heavy target particle is encoded in various Wilson coefficients, which essentially represent various
multipole moments. They can be in principle evaluated  by nonperturbative means, or
can be determined by the bottom-up approach, {\it e.g.}, extracted from low-energy Rutherford scattering experiments.

In analogy with HQET, we build up an EFT dubbed heavy particle effective theory (HPET), describing a static heavy composite fermionic target particle interacting
with soft photon:
\begin{align}
\mathcal{L}_{\rm HPET}=\bar{h}_v\left(i
D_0+c_2\dfrac{\mathbf{D}^2}{2M}+c_Fe\dfrac{\boldsymbol{\sigma}\cdot\mathbf{B}}{2M}+c_De\dfrac{\left[\boldsymbol{\nabla}\cdot\mathbf{E}\right]}{8M^2}+
ic_Se\dfrac{\mathbf{\sigma}\cdot\left(\mathbf{D}\times\mathbf{E}-\mathbf{E}\times\mathbf{D}\right)}{8M^2}\right) h_v + {\mathcal O}(1/M^3),
\label{HPET:lagrangian}
\end{align}
where we have truncated the effective lagrangian through order $1/M^2$.
$h_v$ represents the heavy target HPET field, with the label velocity $v^\mu=(1,{\bf 0})$.
$D^\mu=\partial^\mu+iZeA^\mu$ signifies the covariant derivative,
$\mathbf{E}$ and $\mathbf{B}$ denote
the electric and magnetic field,
The coefficient $c_2=1$ is a rigorous consequence of Lorentz symmetry.
The $c_F$, $c_D$ and $c_S$-related
terms are often referred to as Fermi, Darwin and spin-orbital terms.
The organization of the HPET lagrangian is
governed by powers of $|{\bf q}|/M$, with ${\bf q}$ signifying the photon momentum.

\subsection{HPET description of massless spin-1/2 projectile hitting static spin-${1\over 2}$ target}
\label{HPET:perspective}

In contrast to Fig.~\ref{fig:feynman diag}, up to order $1/M^2$ there arise five Feynman diagrams
in the context of HPET for the tree-level process $e(k) N(p)\to e(k^\prime) N(p^\prime)$.
The corresponding amplitude reads
\bqa
  \mathcal{M}_{\rm HPET} &= &
  -\sqrt{1+c_2\dfrac{\mathbf{p}'^2}{2M^2}}
  \dfrac{e^2}{q^2}
  \bigg\{-Z\bar{u}_\text{NR}u_\text{NR}\bar{u}(k')\gamma^0u(k)
  +\dfrac{c_2Z}{2M}\bar{u}_\text{NR}u_\text{NR}\bar{u}(k')\mathbf{p}'\cdot\boldsymbol{\gamma} u(k)
\nn\\
& & - \dfrac{c_F}{4M}\bar{u}_\text{NR}\left[\slashed q,\gamma^\mu\right]u_\text{NR}\bar{u}(k')\gamma_\mu u(k)
  -\dfrac{c_Dq^2}{8M^2}\bar{u}_\text{NR}u_\text{NR}\bar{u}(k')\gamma^0 u(k)\bigg\}
\nn\\
& = &
  \dfrac{e^2}{q^2}
  \bigg\{ Z\bar{u}_\text{NR}u_\text{NR}\bar{u}(k')\gamma^0u(k)
 + \dfrac{c_F}{4M}\bar{u}_\text{NR}\left[\slashed q,\gamma^\mu\right]u_\text{NR}\bar{u}(k')\gamma_\mu u(k)
  +\dfrac{c_Dq^2}{8M^2}\bar{u}_\text{NR}u_\text{NR}\bar{u}(k')\gamma^0 u(k)
  \bigg\},
\label{massless:projectile:EFT:amplitude}
\eqa
with $u_\text{NR}$ denotes the HPET spinor wave function that satisfies $\slashed v u_\text{NR}(v,s)=u_\text{NR}(v,s)$ and
$\bar{u}_\text{NR}(v,s)u_\text{NR}(v,s)=2v^0$. Notice that final expression does not depend on $c_2$.
Also note the contribution from the $c_S$ term is proportional to $q^0\sim \mathbf{k}^2/M$, hence needs not be considered
at the prescribed accuracy of $1/M^2$.

Squaring the amplitude in \eqref{massless:projectile:EFT:amplitude} and summing/averaging over polarizations, we obtain
the differential unpolarized cross section obtained from the HPET side:
\begin{align}
\dfrac{\mathrm{d}\sigma}{\mathrm{d}\cos\theta}\bigg|_{\rm EFT} = &\dfrac{\pi\alpha^2Z^2\cos^2\frac{\theta}{2}}
{2 {\bf k}^2 \sin^4\frac{\theta}{2}}-\frac{\pi\alpha^2Z^2\cos^2\frac{\theta}{2}}{M |{\bf k}| \sin^2\frac{\theta}{2}}
\nn\\
- &\dfrac{\pi\alpha^2}{8M^2\sin^2\frac{\theta}{2}}\left[Z^2\left(\cos 2\theta-1\right)
+c_D Z\left(\cos\theta+1\right)+c_F^2\left(\cos\theta-3\right)\right],
\label{eq:cross:section-HPET}
\end{align}
Note the LO and NLO terms are only sensitive to the total charge $Ze$ of the target particle, and do not depend on any nontrivial Wilson coefficients (hence
insensitive to the target's internal structure).
This indicates that these two terms are solely dictated by the leading HPET lagrangian.
Since the leading HPET lagrangian possesses heavy particle spin symmetry, the spin degree freedom
is completely decoupled at the lowest order in $1/M$ expansion.
From a technical perspective, the first two terms
arise solely from expanding
the squared LO amplitude in \eqref{massless:projectile:EFT:amplitude} as well as
expanding factor $|{\bf k}^\prime|/|{\bf k}|$ in the phase space measure in \eqref{def:diff:cross:section:massless:lab:frame}.

At first sight, one may worry that the interference between the $c_F$ term
and the LO amplitude would nominally generate a ${\cal O}(1/M)$ correction, thus break the universality at NLO.
A closer examination reveals that this contribution actually vanishes
after summing over polarizations. This cancelation is anticipated to persist for other species of spinning target particle.

The first two terms in \eqref{eq:cross:section-HPET} are indeed identical to
the universal behaviors revealed in \eqref{eq: cross section-NLO}. The effective field theory approach helps us
to better understand why they are independent of the
species of the composite target particle, though our HPET lagrangian only specializes to the spin-$1/2$ target.

At NNLO in $1/M$ expansion, there emerge three terms which stem from different sources.
The first term clearly stems from expanding the squared LO amplitude in combination with the phase space expansion,
which is of the same origin as the LO and NLO contributions, and is anticipated to be universal.
The second term comes from the interference between the $c_D$ term and the LO amplitude, and
the last term stems from the square of the $c_F$ term. Concretely speaking
the last two terms depend on the composite target particle's
charge radius and magnetic dipole.
Interestingly, the $c_F$ term can be identified with the $F^2_{2,0}(\cos\theta-3)$ term in
\eqref{eq: cross section-NNLO}.
As discussed in the paragraph after \eqref{eq: cross section-NNLO}, the coefficient of this term
may depend on the target spin in a specific manner.

To verify that the EFT amplitude does capture the correct soft behavior, we can perform the
heavy target mass expansion from the full QED amplitude in \eqref{eq: amplitude}:
\begin{align}
  {\cal M}_{\rm QED} = & \frac{e^2}{q^2}\bar{u}(k^\prime)\gamma^\mu u(k) \bar{u}(p',\lambda')\left[2P^\mu F_{1,0}\left(\dfrac{q^2}{M^2}\right)+\mathrm{i}\sigma^{\mu\nu}q_\nu F_{2,0}\left(\dfrac{q^2}{M^2}\right)\right]u(p,\lambda)
\nn\\
  =&\frac{e^2}{q^2}\bar{u}\gamma^\mu u\bar{u}_\text{NR}^{\lambda'}\sqrt{\dfrac{p'^0}{M}}\left(1-\dfrac{\mathbf{p}'\cdot\boldsymbol{\gamma}}{2M}-\dfrac{\mathbf{p}'^2}{8M^2}\right)\left[2P^\mu \left(F_{1,0}+F'_{1,0}\dfrac{q^2}{M^2}\right)+\mathrm{i}\sigma^{\mu\nu}q_\nu F_{2,0}\right]u_\text{NR}^\lambda
  \nn\\
    =&\frac{2Me^2}{q^2}\left[
    Z\bar{u}\gamma^0 u\bar{u}_\text{NR}^{\lambda'} u_\text{NR}^\lambda
    +\dfrac{F_{2,0}}{4M}\bar{u}\gamma_\mu u\bar{u}_\text{NR}^{\lambda'}\left[\slashed q, \gamma^\mu\right] u_\text{NR}^\lambda
    +\dfrac{q^2\left(8F'_{1,0}-F_{1,0} + 2F_{2,0}\right)}{8M^2}\bar{u}\gamma^0 u\bar{u}_\text{NR}^{\lambda'}u_\text{NR}^\lambda\right]
\label{massless:projectile:QED:amplitude}
\end{align}
where we have not only expanded the form factor $F_{1,0}$ to the first order in $q^2/M^2$, but also expanded the Dirac spinor using
$u(p')=\sqrt{p^{\prime 0} \over M}\left(1-\dfrac{\mathbf{p}'\cdot\boldsymbol{\gamma}}{2M}-\dfrac{\mathbf{p}'^2}{8M^2}\right) u_\text{NR}+\mathcal{O}(1/M^3)$.

Note the HPET amplitude assumes nonrelativistic normalization for target particle, therefore one needs to include an overall factor $2M$
prior to comparing it with the full QED amplitude.
By equating \eqref{massless:projectile:EFT:amplitude} and \eqref{massless:projectile:QED:amplitude},
we are able to identify the relation between the Wilson coefficients in HPET and
the electromagnetic form factors near the zero-momentum transfer:
\begin{subequations}
\begin{align}
c_F=&F_{2,0},\\
c_D=&2F_{2,0}+8F'_{1,0}-F_{1,0},
\end{align}
\label{eq: HPET-Wilson coefficient}
\end{subequations}
which are identical to those relations obtained for the structureless quark in HQET~\cite{Manohar:1997qy}.

Substituting the relations \eqref{eq: HPET-Wilson coefficient} into
\eqref{eq:cross:section-HPET}, we fully reproduce the NNLO contribution for a heavy spin-$1/2$ target,
as recorded in (\ref{eq: cross section-NNLO-1/2}).

\subsection{NRQED+HPET description of slowly-moving spin-1/2 projectile hitting static spin-${1\over 2}$ target}
\label{NRQED:HPET:perspective}

Next we turn to the EFT approach to understand the second type of Rutherford scattering, a light non-relativistic  particle hits a static heavy composite target.
To be specific, we specialize to a spin-$1/2$ structureless projectile and a spin-$1/2$ target particle. The treatment of the static composite
fermionic target is identical as the section~\ref{HPET:perspective}. It is natural to apply the nonrelativistic QED (NRQED)~\cite{Caswell:1985ui}
to describe the incident slowly-moving electron.

Up to relative order $v^2$, the electron sector of the NRQED Lagrangian reads
\beq
\mathcal{L}_\text{NRQED}=\psi^\dagger\bigg[iD^0 + d_2\dfrac{\mathbf{D}^2}{2 m}+d_4\dfrac{\mathbf{D}^4}{8 m^3}+d_Fe
\dfrac{\boldsymbol{\sigma}\cdot\mathbf{B}}{2 m}+d_De\dfrac{\left[\boldsymbol{\nabla}\cdot\mathbf{E}\right]}{8 m^2}
+ id_Se\dfrac{\boldsymbol{\sigma}\cdot\left(\mathbf{D}\times\mathbf{E}-\mathbf{E}\times\mathbf{D}\right)}{8 m^2}\bigg]\psi,
\label{NRQED:lagrangian}
\eeq
where $\psi$ denotes a Pauli spinor field that annihilates a nonrelativistic electron.
$d_2=d_4=1$ is a rigorous consequence of Lorentz invariance. The $d_4$ term, together with the $d_F$, $d_D$ and $d_S$ terms (referred to as the
Fermi, Darwin and spin-orbital terms), represent the $O(v^2)$
corrections to NRQED lagrangian. At tree level, the Wilson coefficients $d_F = d_D= d_S = 1$.

Our starting point is the HPET lagrangian \eqref{HPET:lagrangian} and the NRQED lagrangian \eqref{NRQED:lagrangian}. It is convenient to
work in Coulomb gauge. Up to ${\cal O}(v^2/M^2)$, the relevant tree-level EFT amplitude for $e N\to e N$ reads
\begin{align}
\label{NRQED+HPET:amplitude}
  \mathcal{M}_{\rm EFT}= & \dfrac{e^2}{\mathbf{q}^2}\xi^\dagger \left[1+\dfrac{d_2}{4m^2}(\mathbf{k}^2+\mathbf{k}'^2)-\dfrac{d_D}{8m^2}
  |\mathbf{k}'-\mathbf{k}|^2-\dfrac{i d_S}{4m^2}\boldsymbol{\sigma}\cdot(\mathbf{k}\times \mathbf{k}')\right] \xi \, \bar{u}_\text{NR}^{\lambda'}
    \Big[-Z+\dfrac{\left(c_D-2c_2Z\right)\mathbf{p}'^2}{8M^2}
    \Big]u_\text{NR}^\lambda
\\
- & \dfrac{1}{q^2}\left(\delta^{ij}-\dfrac{q^iq^j}{\mathbf{q}^2}\right)\xi^\dagger \bigg\{
      (k^i+k^{'i})
      \left[\dfrac{d_2}{2m}+\dfrac{d_2^2-d_4}{8m^3}(\mathbf{k}^2+\mathbf{k}'^2)\right]
      +\dfrac{i d_F}{2m}\left(1+ d_2 {\mathbf{k}^2+\mathbf{k}'^2 \over 4 m^2}\right)\left[\boldsymbol{\sigma}\times(\mathbf{k}'-\mathbf{k})\right]^i
\nn\\
- & \dfrac{d_D}{16m^3}(k'^i-{k}^i)(\mathbf{k}'^2-\mathbf{k}^2) -\dfrac{i d_S}{16m^3}(\mathbf{k}'^2-\mathbf{k}^2)\left[\boldsymbol{\sigma}\times(\mathbf{k}+\mathbf{k}')\right]^i\bigg\}\xi \,  \bar{u}_\text{NR}^{\lambda'}
    \Big[-\dfrac{c_2Z}{2M}{p}'^j-i\dfrac{c_F}{2M}\sigma^{jl}{p}'^l
    \Big]u_\text{NR}^\lambda,
\nn
\end{align}
where the first line represents the temporal photon exchange, and the remaining lines represent the transverse photon exchange. $\xi$ denotes
the two-component spinor wave function.

After some simplification, \eqref{NRQED+HPET:amplitude} reduces to
\begin{align}
  \mathcal{M}_{\rm EFT}
    =&\dfrac{e^2}{\mathbf{q}^2}\Big[-Z+\dfrac{\left(c_D-2c_2Z\right)\mathbf{p}'^2}{8M^2}
    \Big]
   \xi^\dagger\left[1+\dfrac{d_2}{4m^2}\left(\mathbf{k}^2+\mathbf{k}'^2\right)-\dfrac{d_D}{8m^2}|\mathbf{k}'-\mathbf{k}|^2-\dfrac{\mathrm{i} d_S}{4m^2}\boldsymbol{\sigma}\cdot(\mathbf{k}\times \mathbf{k}')\right]\xi\bar{u}_\text{NR}^{\lambda'}
    u_\text{NR}^\lambda
\nn\\
 - &\dfrac{c_Fe^2}{4M\mathbf{q}^2}\xi^\dagger\bigg\{
      (k^i+{k}'^i)
      \left[\dfrac{d_2}{2m}+\dfrac{d_2^2-d_4}{8m^3}(\mathbf{k}^2+\mathbf{k}'^2)\right]
      +\dfrac{\mathrm{i} d_F}{2m}\left[\boldsymbol{\sigma}\times(\mathbf{k}'-\mathbf{k})\right]^i-\dfrac{\mathrm{i} d_S}{16m^3}(\mathbf{k}'^2-\mathbf{k}^2)\left[\boldsymbol{\sigma}\times(\mathbf{k}'-\mathbf{k})\right]^i\bigg\}\xi
\nn\\
 \times   &
 \bar{u}_\text{NR}^{\lambda'}
    \Big[\gamma^i,\boldsymbol{\gamma}\cdot \mathbf{q}
    \Big]u_\text{NR}^\lambda.
\label{NRQED+HPET:amplitude:final}
\end{align}

Squaring the amplitude in \eqref{NRQED+HPET:amplitude:final}, summing/averaging over various spins,
we obtain the differential unpolarized Rutherford scattering cross section in the context of EFT:
\begin{align}
  \dfrac{\mathrm{d}\sigma}{\mathrm{d}\cos\theta}
\bigg|_{\rm EFT} =& \: \frac{\pi  \alpha ^2 m^2Z^2}{2\mathbf{k}^4 \sin^4\frac{\theta}{2}}
  -\frac{\pi  \alpha ^2  m^4Z^2}{M^2\mathbf{k}^4}
  +\dfrac{\pi\alpha^2Z}{2\mathbf{k}^{2}\sin^2\frac{\theta}{2}}
  \bigg\{\dfrac{Z\left(d_D\cos\theta-d_D+2\right)}{2\sin^2\frac{\theta}{2}}
\nn\\
-& {m\over M} (d_D\cos\theta-d_D+2)
-{m^2\over 2 M^2} \left[ Z(2+d_D-4c_2)+Z\cos\theta (2-d_D) + 2c_D\right]\bigg\},
\label{NRQED+HPET:unpol:cross:section}
\end{align}
The $d_S$ term in \eqref{NRQED+HPET:amplitude:final} does not contribute to the squared amplitude since its interference with LO amplitude in velocity expansion
only contains a single Pauli matrix, hence vanishes upon summing over polarization.

Substituting $c_2=d_D=1$ in \eqref{NRQED+HPET:unpol:cross:section}, and utilizing the relations given in \eqref{eq: HPET-Wilson coefficient},
we exactly reproduce (\ref{eq: NR-LO}) which encodes the LO and NLO terms in heavy target expansion, as well as
\eqref{eq:NR-NNLO:spin-half} which encapsulates the NNLO term.
(\ref{eq:NR-NNLO}) indicates that the $Z\cos\theta$ and $F^\prime_{1,0}$ terms in NNLO correction are universal, {\it e.g.},
independent of the target spin. This may indicate the structures such as $Z\cos\theta (2-d_D) + 2c_D$  may arise ubiquitously
in an EFT calculation for the heavy target other than spin-$1/2$ fermion.

To verify that the EFT amplitude indeed reproduces the correct soft behavior, we conduct both nonrelativistic and heavy target mass expansion
from the full QED amplitude in \eqref{eq: amplitude}.

Working again in Coulomb gauge, and employing the
following the relation between the relativistic electron spinor and nonrelativistic
electron spinor:
\begin{align}
  u(k)=\dfrac{1}{\sqrt{k^0+m}}
  \begin{pmatrix}
    \left(k^0+m\right)\xi \nn\\
    \mathbf{k}\cdot\boldsymbol{\sigma}\xi
  \end{pmatrix}
,\quad \bar{u}(k)=\dfrac{1}{\sqrt{k^0+m}}
\begin{pmatrix}
\left(k^0+m\right)\xi^\dagger&
-\xi^\dagger\mathbf{k}\cdot\boldsymbol{\sigma}
\end{pmatrix},
\end{align}
we expand the full QED amplitude through ${\cal O}(v^2/M^2)$:
\begin{align}
\label{NR:Expanded:QED:ampl:spin:half:target}
  \mathcal{M}_{\rm QED} = &
  -\frac{e^2}{\mathbf{q}^2}\bar{u}(k^\prime)\gamma^0 u(k) \bar{u}_\text{NR}^{\lambda'}\left[
    2ZP^0
    +\dfrac{P^0}{4M^2}\left(8F'_{1,0}q^2+Z{\mathbf{q}^2}\right)
    +\mathrm{i}\dfrac{\mathbf{p}'\cdot\boldsymbol{\gamma}}{2M}\sigma^{0i}q_i F_{2,0}\right]u_\text{NR}^\lambda
\\
    &-\frac{e^2}{\mathbf{q}^2}\left(\delta_{ij}-\dfrac{q^iq^j}{\mathbf{q}^2}\right)\bar{u}\gamma^i u\bar{u}_\text{NR}^{\lambda'}\left[
      2ZP^j
      +\dfrac{1}{2}\left[\slashed q, \gamma^j\right] F_{2,0}
      +\dfrac{P^j}{4M^2}\left(8F'_{1,0}q^2+Z{\mathbf{q}^2}\right)
      \right]u_\text{NR}^\lambda
\nn\\
=&\frac{2Me^2}{\mathbf{q}^2}2m\bigg[
          -Z
          +\dfrac{\mathbf{p}'^2}{8M^2}\bigg(2F_{2,0}+8F'_{1,0}-3Z\bigg)
          \bigg]\xi^\dagger\left[
          1
        +\dfrac{\left|\mathbf{k}+\mathbf{k}'\right|^2}{8m^2}
        -\dfrac{\mathrm{i}}{4m^2} \boldsymbol{\sigma}\cdot\left(\mathbf{k}\times\mathbf{k}'\right)\right]\xi
        \bar{u}_\text{NR}^{\lambda'}u_\text{NR}^\lambda
          \nn\\
          &-\frac{F_{2,0}e^2}{2\mathbf{q}^2}2m\xi^\dagger\bigg\{
            \dfrac{1}{2m}\left({k}^i+{k}'^i\right)+\dfrac{i}{2m}\left[\boldsymbol{\sigma}\times\left(\mathbf{k}'-\mathbf{k}\right)\right]^i-\dfrac{i}{16m^3}\left(\mathbf{k}'^2-\mathbf{k}^2\right)\left[\boldsymbol{\sigma}\times\left(\mathbf{k}'-\mathbf{k}\right)\right]^i
          \bigg\}\xi
          \bar{u}_\text{NR}^{\lambda'}\left[\gamma^i,\boldsymbol{\gamma}\cdot\mathbf{q}
            \right]u_\text{NR}^\lambda.
\nn\end{align}

After including the normalization factor $(2M)(2 m)$, employing the relations for the heavy target Wilson coefficients
in \eqref{eq: HPET-Wilson coefficient}, and taking $d_2=d_4=d_F=d_D=d_S=1$,
one finds that the EFT amplitude
\eqref{NRQED+HPET:amplitude:final} exactly agrees with the full QED amplitude
\eqref{NR:Expanded:QED:ampl:spin:half:target}.

\section{Summary}
\label{sec:summary}

In this work, we have conducted a comprehensive study of the soft behavior of the tree-level
Rutherford scattering process. We have considered two classes of Rutherford scattering experiments,
a low-energy point-like massless projectile ({\it e.g.}, a spin-${1\over 2}$ or spin-$0$ electron)
bombs a static massive composite spinning target particle ({\it e.g.}, atomic nucleus), and a slowly-moving light structureless projectile hits
a static heavy composite spinning target. We have considered various composite target particle with spin up to 2.

The soft limits of the unpolarized cross sections in the laboratory frame in both cases exhibit some universal pattern.
For the former type of Rutherford scattering process, given a specific projectile, the first two terms in the differential cross section are universal upon heavy target mass
expansion, while the universality starts to break down at NNLO. 
Nevertheless, many terms at NNLO still remain to be spin-independent or have some definite spin-dependence pattern.
For the latter type, we have to perform both nonrelativistic and heavy target mass expansion to infer the correct soft limit. 
At the lowest order in projectile
velocity expansion yet to all orders in $1/M$ expansion, the differential cross section has a universal form (insensitive to the projectile spin).
At NLO in velocity expansion, the first two terms in the differential cross section in $1/M$ expansion are still universal.
The ${\cal O}(v^2/M^2)$ piece starts to partially violate the universality. Despite this, some terms at this order still remain to be target spin
independent.

It is of special interest that the $F_{2,0}$ term at ${\cal O}(1/M^2)$ (or ${\cal O}(v^2/M^2)$ for second type of Rutherford scattering) 
seems to reflect some peculiar spin-statistics feature. 
Its coefficient remains to be one constant for fermionic target, while another constant for bosonic target.
It is interesting to verify this observation by investigating the target particle with even higher spin.

We have also attempted to apply effective field theory approach to understand the soft pattern of the Rutherford scattering cross sections, taking
the target particle as a composite Dirac fermion for concreteness. 
Some useful insight is gained from the EFT perspective.

\begin{acknowledgments}
We are grateful to the useful discussions with Zhewen Mo and Jichen Pan.
This work is supported in part by the National Natural Science Foundation of China under Grants No. 11925506 and
No. 12070131001 (CRC110 by DFG and NSFC).
\end{acknowledgments}

\appendix

\section{Rutherford scattering with massless spinless projectile}

We can repeat our investigation in Section~\ref{mass:0:fermion:projectile} by replacing the projectile to be a massless spin-0 electron,
which is described by scalar QED. The electromagnetic vertex involving scalar electron is simply given by
\begin{align}
  &\langle e (k')\vert J^\mu\vert e(k)\rangle=-\left(k^\mu+k'^\mu\right).
\end{align}

Upon heavy target mass expansion, we again observe that the unpolarized cross sections exhibit some universal feature.
Concretely speaking, the LO and NLO pieces in $1/M$ expansion are independent of the target particle spin:
\begin{align}
  &\dfrac{\mathrm{d}\sigma}{\mathrm{d}\cos\theta}=\dfrac{\pi\alpha^2Z^2}{2{\bf k}^2\sin^4\left(\frac{\theta}{2}\right)}
  -\frac{\pi\alpha^2Z^2}{M |{\bf k}|\sin ^2\left(\frac{\theta}{2}\right)}+{\cal O}\left({1\over M^2}\right).
\end{align}
The universality becomes partially violated at NNLO.
For various target particle, the NNLO contributions are
\begin{subequations}
  \begin{align}
    \left(\dfrac{\mathrm{d}\sigma}{\mathrm{d}\cos\theta}\right)_\text{NNLO}^{s=0}=&-\dfrac{4\pi\alpha^2}
    {M^2\sin^2\frac{\theta}{2}}{\left(F'_{1,0}Z+\dfrac{5}{16}Z^2\cos{\theta}-\dfrac{5}{16}Z^2\right)},\\
    \left(\dfrac{\mathrm{d}\sigma}{\mathrm{d}\cos\theta}\right)_\text{NNLO}^{s=\frac{1}{2}}
    =&-\dfrac{4\pi\alpha^2}{M^2\sin^2\frac{\theta}{2}}
    \bigg[F'_{1,0}Z-\frac{1}{16}F^2_{2,0}(\cos\theta+1)+\frac{1}{4}F_{2,0}Z\nn\\
    &+\frac{5}{16}Z^2\cos \theta-\frac{7}{16}Z^2\bigg],
  \label{spinless:projectile:target:spin:half}
    \\
    \left(\dfrac{\mathrm{d}\sigma}{\mathrm{d}\cos\theta}\right)_\text{NNLO}^{s=1}=&-\dfrac{4\pi\alpha^2}{M^2\sin^2\frac{\theta}{2}}
    \bigg[F'_{1,0}Z-\dfrac{1}{24}F^2_{2,0}(\cos\theta+1)+\frac{1}{6}F_{2,0}Z\nn\\
    &+\frac{5}{16}Z^2\cos\theta-\frac{1}{6}F_{1,1}Z-\frac{23}{48}Z^2\bigg],\\
    \left(\dfrac{\mathrm{d}\sigma}{\mathrm{d}\cos\theta}\right)_\text{NNLO}^{s=\frac{3}{2}}
    =&-\dfrac{4\pi\alpha^2}{M^2\sin^2\frac{\theta}{2}}\bigg[
      F'_{1,0}Z-\frac{5}{144}F^2_{2,0}(\cos\theta+1)+\frac{1}{4}F_{2,0}Z\nn\\
    &+\frac{5}{16}Z^2\cos \theta-\frac{1}{6}F_{1,1}Z-\frac{5}{144}F_{2,0}^2-\frac{29}{48}Z^2\bigg],\\
    \left(\dfrac{\mathrm{d}\sigma}{\mathrm{d}\cos\theta}\right)_\text{NNLO}^{s=2}=&-\dfrac{4\pi\alpha^2}{M^2\sin^2\frac{\theta}{2}}\bigg[F'_{1,0}Z-\frac{1}{32}F^2_{2,0}(\cos \theta+1)+\frac{1}{6} F_{2,0}Z\nn\\
    &+\frac{5}{16}Z^2\cos \theta-\frac{1}{6}F_{1,1}Z-\frac{31}{48}Z^2\bigg].
\end{align}
\end{subequations}

Similar to the pattern indicated in \eqref{eq: cross section-NNLO} for a massless spin-${1\over 2}$ projectile,
we observe that $F'_{1,0}Z$, $Z^2\cos\theta$ and $F_{1,1}Z$ terms are independent of the target spin.
The $F'_{1,0}Z$ and $Z^2\cos\theta$ terms actually have the same origin of the LO and NLO cross sections,
which correspond to different terms in Taylor expansion of $F^2_{1,0}(q^2/M^2)$ in the squared LO amplitude and phase space measure.
The coefficient of the $F_{2,0}Z$ term seems to reflect the spin-statistic characteristic of the target particle.
For fermions, the coefficient is $1/4$, while for bosons $1/6$.

Although the coefficients of $F^2_{2,0}(\cos\theta+1)$ inside the square bracket
explicitly depend on the target spin $s$,  they seem to be expressed as
$-{1+s\over 48s}$, at least for $s=1/2,1,3/2,2$. It will be interesting to see whether this parameterization
persists for an arbitrary $s$ or not.

Analogous to what is done in Section~\ref{mass:0:fermion:projectile}, for a spin-$1/2$ composite target particle,
the HPET-based calculation yields the following unpolarized cross section:
\beq
  \dfrac{\mathrm{d}\sigma}{\mathrm{d}\cos\theta}=\dfrac{\pi\alpha^2Z^2}{2{\bf k}^2\sin^4\frac{\theta}{2}}
  -\frac{\pi\alpha^2Z^2}{M |{\bf k}|\sin^2\frac{\theta}{2}}+\dfrac{\pi\alpha^2}{4M^2\sin^2\frac{\theta}{2}}
  \Big[-2c_DZ+c_F^2(\cos\theta+1)+5Z^2(1-\cos\theta)\Big].
\eeq
Reassuringly, this EFT result exactly agrees with what is obtained from \eqref{spinless:projectile:target:spin:half}.

\section{Rutherford scattering with nonrelativistic spinless projectile}

We can repeat our investigation in Section~\ref{light:NR:spin:half:projectile} by replacing the projectile to be a light slowly-moving spinless electron.
At lowest order in electron velocity, yet to all orders in $1/M$, the resulting unpolarized cross section is identical to \eqref{eq: NR-LO}£¬which was
obtain for a spin-${1\over 2}$ projectile.
This is well anticipated, since the spin degree of freedom decouples in the nonrelativistic limit.

At relative order-$v^2$, after the heavy target mass expansion, the differential unpolarized cross section becomes particularly simple:
\begin{align}
  \left(\dfrac{\mathrm{d}\sigma}{\mathrm{d}\cos\theta}\right)_{(v^2)}=&
  \dfrac{\pi\alpha^2}{2\mathbf{k}^2\sin^2\frac{\theta}{2}}
  \left[\frac{Z^2}{\sin^2\frac{\theta}{2}}-\frac{2mZ^2}{M}-\dfrac{4m^2Z}{M^2}\tilde{f}^s_\text{NNLO}+\mathcal{O}\left(\dfrac{1}{M^3}\right)\right],
\end{align}
where
\begin{subequations}
  \begin{align}
\tilde{f}^{s=0}_\text{NNLO}=&2F'_{1,0}
+\dfrac{Z}{4}\cos\theta
-\frac{Z}{4},
\\
\tilde{f}^{s=1/2}_\text{NNLO}=&2F'_{1,0}+\dfrac{1}{2}F_{2,0}
+\dfrac{Z}{4}\cos\theta
-\frac{Z}{2},
\label{spin:0:projectile:spin:half:target:soft}\\
\tilde{f}^{s=1}_\text{NNLO}=&2F'_{1,0}+\dfrac{1}{3}F_{2,0}
-\dfrac{1}{3}F_{1,1}
+\dfrac{Z}{4}\cos\theta
-\frac{7Z }{12}\\
\tilde{f}^{s=3/2}_\text{NNLO}=&2F'_{1,0}+\dfrac{1}{2}F_{2,0}
-\dfrac{1}{3}F_{1,1}
+\dfrac{Z}{4}\cos\theta
-\frac{5Z}{6},
\\
\tilde{f}^{s=2}_\text{NNLO}=&2F'_{1,0}+\dfrac{1}{3}F_{2,0}
-\dfrac{1}{3}F_{1,1}
+\dfrac{Z}{4}\cos\theta
-\frac{11Z}{12}.
\end{align}
\end{subequations}
Clearly the ${\cal O}(v^2/M^n)$ ($n=0,1$) terms remain universal.
At ${\cal O}(v^2/M^2)$, the universality becomes partially violated. However,
the $F^\prime_{1,0}$, $F_{1,1}$ and $Z\cos\theta$ terms still do not depend on the
target particle spin. The coefficient of $F_{2,0}$ seems to reflect the spin-statistic characteristic of the target particle.
For fermions, the coefficient is $1/2$, while for bosons $1/3$.

Similar to Section~\ref{NRQED:HPET:perspective}, we can combine NRQED and HPET to study the soft behavior of this type of Rutherford scattering.
Since the incident electron is assumed to be spinless, it is natural to work with scalar NRQED plus HPET.
Up to the relative order-$v^2$, the scalar NRQED lagrangian reads
\begin{align}
    \mathcal{L}_\text{sNRQED}=& Q^\dagger\left(iD^0+d_2\dfrac{\mathbf{D}^2}{2m}+d_4\dfrac{\mathbf{D}^4}{8m^3}\right)Q,
\end{align}
with $Q$ signifying the field that annihilates a nonrelativistic scalar electron. Again $d_2=d_4=1$ is a rigorous consequence of
Lorentz symmetry.

Based on the scalar NRQED and HPET, we are able to obtain the following
unpolarized Rutherford cross section, which is accurate to the relative
order-$v^2/M^2$:
\begin{align}
  \dfrac{\mathrm{d}\sigma}{\mathrm{d}\cos\theta}=&\frac{\pi  \alpha^2 m^2Z^2}{2\mathbf{k}^4 \sin^4\frac{\theta}{2}}
  -\frac{\pi  \alpha ^2  m^4Z^2}{M^2\mathbf{k}^4}
  +\dfrac{Z\pi\alpha^2}{2\mathbf{k}^2\sin^2\frac{\theta}{2}}\bigg[\dfrac{Z}{\sin^2\frac{\theta}{2}}
-\dfrac{2mZ}{M}
-\frac{m^2\left(-2c_2Z+c_D+Z\cos\theta+Z\right)}{M^2}\bigg].
\end{align}
Reassuringly, this EFT result exactly reproduce the soft limit obtained from the full QED,  \eqref{spin:0:projectile:spin:half:target:soft}.

One can further verify that the EFT amplitude indeed reproduces the correct soft behavior, which is deduced by
conducting both nonrelativistic and heavy target mass expansion from the full QED amplitude in \eqref{eq: amplitude}.
Working again in Coulomb gauge, and using the following electromagnetic matrix element involving spinless electron
\begin{align}
  \langle k'\vert J^0\vert k\rangle= &-1,\\
  \langle k'\vert \mathbf{J}\vert k\rangle=&(\mathbf{k}+\mathbf{k}')\left[\dfrac{d_2}{2m}-\dfrac{d_4}{8m^3}(\mathbf{k}^2+|\mathbf{k}'|^2)\right],
\end{align}
one can readily obtain the expanded Rutherford amplitude through order-$v^2/M^2$, which is indeed compatible with the EFT amplitude.


\end{document}